\documentclass[12pt,english,aps,prb,preprint,superscriptaddress,showpacs,showkeys]{article}
\usepackage[T1]{fontenc}
\usepackage[latin9]{inputenc}
\usepackage{textcomp}
\usepackage{amsmath}
\usepackage{amssymb}
\usepackage{graphicx}
\usepackage{subscript}

\makeatletter

\providecommand{\tabularnewline}{\\}

\newcommand{\lyxaddress}[1]{
\par {\raggedright #1
\vspace{1.4em}
\noindent\par}
}

\makeatletter

\usepackage{amsfonts}

\usepackage{times}
\usepackage{babel}
\usepackage[super,comma,sort&compress]{natbib}

\usepackage{epstopdf}

\makeatletter

 \usepackage{epstopdf}
    \usepackage{times}
\makeatother

\makeatother

\usepackage{babel}
\begin{document}

\title{Structural phase transformations in metallic grain boundaries}

\author{Timofey Frolov$^{1\star}$, \enskip{}David L. Olmsted$^{1}$,\enskip{}Mark
Asta$^{1}$ and Yuri Mishin$^{2}$}

\maketitle

\lyxaddress{$^{1}$ Department of Materials Science and Engineering, University
of California, Berkeley, California 94720, USA}

\lyxaddress{$^{2}$ School of Physics, Astronomy and Computational Sciences,
George Mason University, Virginia, 22030, USA}

$^{\star}$e-mail: timfrol@berkeley.edu

\bigskip{}

\noindent \begin{center}
\textsf{\textbf{Abstract}}
\par\end{center}

Structural transformations at interfaces are of profound fundamental
interest as complex examples of phase transitions in low-dimensional
systems. Despite decades of extensive research, no compelling evidence
exists for structural transformations in high-angle grain boundaries
in elemental systems. Here we show that the critical impediment to
observations of such phase transformations in atomistic modeling has
been rooted in inadequate simulation methodology. The proposed new
methodology allows variations in atomic density inside the grain boundary
and reveals multiple grain boundary phases with different atomic structures.
Reversible first-order transformations between such phases are observed
by varying temperature or injecting point defects into the boundary
region. Due to the presence of multiple metastable phases, grain boundaries
can absorb significant amounts of point defects created inside the
material by processes such as irradiation. We propose a novel mechanism
of radiation damage healing in metals which may guide further improvements
in radiation resistance of metallic materials through grain boundary
engineering. 

\newpage{}

\noindent \begin{flushleft}
\textsf{\textbf{\large Introduction}}\textsf{\textbf{ }}
\par\end{flushleft}

Phase transformations at interfaces in crystalline materials are of
significant fundamental interest and practical importance due to their
possible impact on the macroscopic mechanical, transport and thermal
properties of polycrystals\cite{Balluffi95,Mishin2010a}. To date,
the studies of interfacial phase transitions have been focused primarily
on ``complexions'' of intergranular thin films in ceramics\cite{Baram08042011,Harmer08042011,Luo23092011}
and pre-melting transitions in metallic alloys\cite{Mishin09c}. Little
is known about grain boundary (GB) transformations in single-component
metals apart from the recently found dislocation pairing transition
in low-angle GBs composed of discrete dislocations\cite{Olmsted2011}.
Despite decades of research, no convincing experimental or simulation
evidence exists for structural transformations in high-angle GBs.
Direct experimental observations of interfacial phase transitions
by high-resolution transmission electron microscopy (HRTEM) are extremely
difficult\cite{Merkle1987,Duscher04,Lacon2004}, and to date evidence
for the possibility of GB transformations has been obtained by less
direct experimental methods. For example, recent measurements of diffusion
in the Cu $\Sigma5\,(310)$ GB ($\Sigma$ being the reciprocal density
of coincident sites\cite{Balluffi95}) revealed a marked change in
the temperature dependence of the diffusivity at about 800-850 K\cite{Divinski2012,Budke1999},
suggesting a possible structural transformation in this boundary.
However, the concurrent molecular dynamics (MD) simulations\cite{Divinski2012}
did not detect any structural change other than gradual accumulation
of disorder and eventually GB pre-melting at high temperatures. 

Due to the difficulty of experimental observations, much of the current
knowledge about GB structures has come from computer simulations.
Most of the MD simulations conducted to date have employed periodic
supercells that did not permit variations in atomic density in the
GB core region, which are possible under experimental conditions.
This unphysical constraint stabilizes one particular GB structure
and prohibits transformations to other structures with different GB
densities. This can explain why transformations in high-angle GBs
have not been found in the previous simulation work.

We propose a more general simulation approach which facilitates variations
in GB density allowing the boundary to assume the most thermodynamically
favorable structure. This approach reveals the existence of multiple
GB phases characterized by not only different structures but also
different densities. This opens the door, for the first time, to direct
observations of phase transitions between GB phases with different
densities in atomistic simulations of metallic systems.

\bigskip{}

\noindent \begin{flushleft}
\textsf{\textbf{\large Results}}\textsf{\textbf{ }}
\par\end{flushleft}

\begin{figure}
\begin{centering}
\includegraphics[height=0.8\textheight]{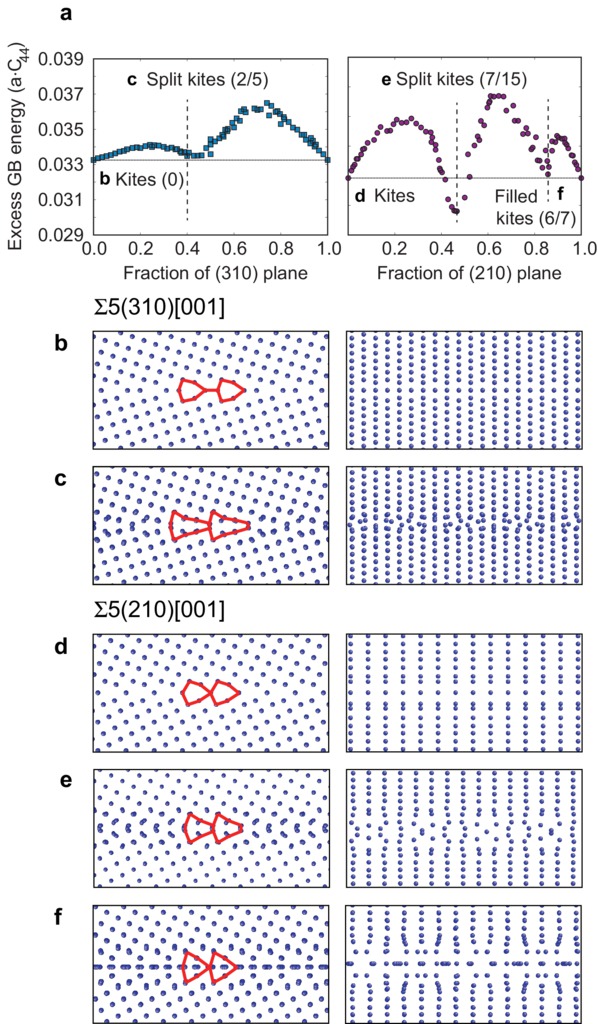}
\par\end{centering}

\caption{\textbf{Calculated structures and energies for the }$\Sigma5\,(310)$
\textbf{and} $\Sigma5\,(210)$\textbf{ GBs at 0K. }Results of GB structure
and energy calculations at 0K using an EAM potential for Cu\cite{Mishin01}.
(\textbf{a}) GB energy versus atomic density (fraction of atomic plane)
for the $\Sigma5\,(310)$ (left) and $\Sigma5\,(210)$ (right) GBs.
The horizontal line marks the energy of the kite structure. The densities
corresponding to local energy minima and the respective GB structures
are indicated. Panels (\textbf{b})-(\textbf{f}) show the GB structures
as viewed parallel to the {[}001{]} tilt axis (left row) and normal
to it (right row). (\textbf{b}) and (\textbf{d}) Normal kites. (\textbf{c})
and (\textbf{e}) Split kites. (\textbf{f}) Filled kites. \label{fig:0K-structures} }
\end{figure}

\noindent \textsf{\textbf{Grain boundary structures at 0 K.}} To demonstrate
this new approach, we studied two representative high-angle, high-energy
$\Sigma5\,(210)$ and $\Sigma5\,(310)$ symmetrical tilt GBs in several
FCC metals with atomic interactions described by embedded-atom method
(EAM)\cite{Daw84} potentials. The choice of the $\Sigma5\,(310)$
GB was motivated by the recent diffusion measurements described above\cite{Divinski2012,Budke1999}.
The long-accepted structure of these $\Sigma5$ boundaries is an array
of kite-shaped structural units (Fig.~\ref{fig:0K-structures}b and
d). This structure is easily obtained by joining two perfect crystallites
along $(210)$ or $(310)$ planes and applying various grain translations
followed by static relaxation. It has been suggested, however, that
in order to sample all possible structures of a GB its density must
be also varied by adding or removing atoms\cite{Chua2010,Alfthan07,Alfthan06,von-Alfthan2006,Phillpot1994,Phillpot1992}.
We applied an algorithm which finds equilibrium GB structures at 0
K by varying the GB density along with grain translations and static
relaxation. In Fig.~\ref{fig:0K-structures} we plot the obtained
energies of the $\Sigma5$ GBs in Cu versus the extra number of atoms
relative to a perfect crystal plane. The minima in this plot correspond
to three different equilibrium structures: (i) the normal kites \emph{b}
and \emph{d}, (ii) split kites \emph{c} and \emph{e} having extra
2/5 of the $(310)$ plane or 7/15 of the $(210)$ plane, respectively;
and (iii) filled kites \emph{f} having extra 6/7 of the $(210)$ plane.
Different cross-sections of the simulation block containing these
structures were tested as illustrated in the Supplementary Figure
S1. 

For each structure corresponding to a local minimum of energy we computed
the excess GB energy $[U]_{N}$, excess volume $[V]_{N}$ and two
components of the GB stress, $\tau_{11}^{N}$ and $\tau_{22}^{N}$,
in the directions parallel and normal to the tilt axis, respectively
\cite{Frolov2012a,Frolov2012b}. The results summarized in Table \ref{tab:310}
clearly demonstrate that the structures are characterized by significantly
different thermodynamic parameters, including even different signs
of the GB stress component $\tau_{22}^{N}$.

To demonstrate the generality of this important finding, similar calculations
were performed with a different EAM Cu potential and for several other
FCC metals including Ag, Au and Ni (see Supplementary Figure S2).
In all cases tested, the structures shown in Fig.~\ref{fig:0K-structures}
represented minima of the GB energy as a function of GB density. Remarkably,
and contrary to the common belief, \emph{none} of the tested potentials
predicts the normal kites to be the lowest-energy structure of the
$\Sigma5\,(210)$ GB! The split kites are energetically more favorable
for this boundary and are only slightly higher in energy than the
normal kites for the $\Sigma5\,(310)$ GB.\bigskip{}

\begin{figure}
\noindent \begin{centering}
\includegraphics[height=0.7\textheight]{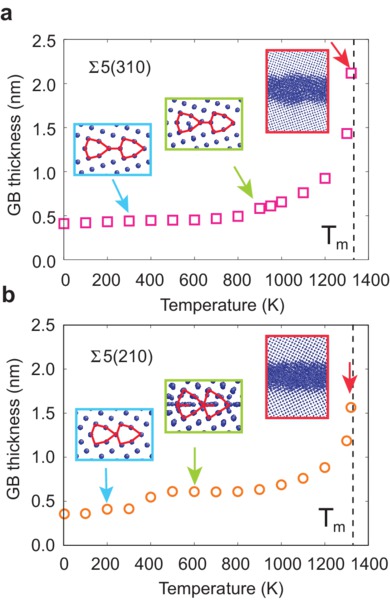}
\par\end{centering}

\caption{\textbf{Effect of temperature on GB structure from MD simulations
with periodic boundary conditions.} The thickness of GBs was evaluated
from the energy density profile across the GB. (\textbf{a}) The $\Sigma5\,(310)$
GB gradually accumulates disorder and pre-melts near the melting point
\textit{T}\textsubscript{m}, whereas (\textbf{b}) the $\Sigma5\,(210)$
GB undergoes a transformation at 400 K before pre-melting near \textit{T}\textsubscript{m}.
\label{fig:pbcT}}

\end{figure}

\noindent \textsf{\textbf{GB phase transformations with temperature.}}
The effect of temperature on the GB structures was studied by MD simulations
with various boundary conditions starting with the normal-kite structure.
First, we tested the traditional methodology with periodic boundary
conditions. In the Cu $\Sigma5\,(310)$ GB, the kites continued to
exist up to $\sim0.75$ of the melting point \textit{T}\textsubscript{m}\textsubscript{}$=1327$
K. At higher temperatures the GB begins to disorder and eventually
pre-melts near \textit{T}\textsubscript{m} (Fig.~\ref{fig:pbcT}a).
This behavior known from previous work\cite{Frolov2012b} cannot explain
the transformation suggested by the diffusion experiments\cite{Divinski2012,Budke1999}.
A more complex behavior was observed in the $\Sigma5\,(210)$ GB.
At about 400 K, the kites were found to transform into a structure
similar to the filled kites (cf.~Fig.~\ref{fig:0K-structures}f).
In Fig.~\ref{fig:pbcT}b a jump in thickness occurs between 300 K
and 400 K due to this change in GB structure. Similar behavior was
found in simulations with other EAM potentials for Cu, Au, Ag and
Ni, suggesting that it represents a generic feature of FCC metals.
However, these simulations impose a constraint on the supply of atoms
into the GB core and do not reveal the full picture of possible structural
changes. 

\begin{figure}
\begin{centering}
\includegraphics[height=0.7\textheight]{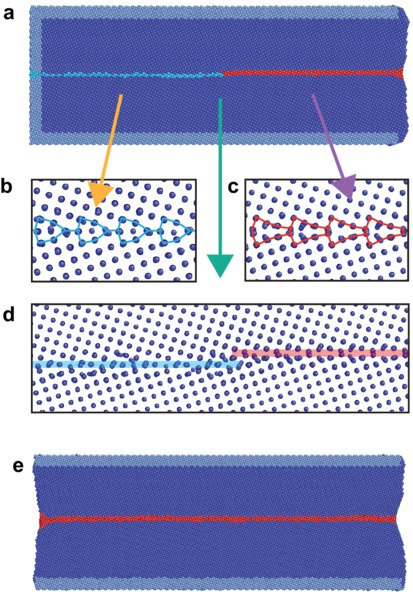}
\par\end{centering}

\caption{\textbf{High temperature transformation of the $\Sigma5\,(310)$ GB.
}(\textbf{a}) MD simulation of the Cu $\Sigma5\,(310)$ GB at 800
K with an open surface allowing variations in GB density. The boundary
undergoes a first order phase transformation with the split-kite structure
nucleating and growing from the surface. The normal-kite (\textbf{b})
and split-kite (\textbf{c}) structures are separated by a 1D phase
boundary accompanied by a step (\textbf{d}). (\textbf{e}) Completely
transformed GB structure in the presence of two open surfaces. \label{fig:310_800_trans} }
\end{figure}

For a more complete investigation, we modified the methodology by
letting the GB terminate at an open surface as shown schematically
in the Supplementary Figure S3. The surface serves as a source or
sink of atoms which may diffuse in or out of the GB to automatically
adjust its density. The MD results are illustrated in Fig.~\ref{fig:310_800_trans}
for an isothermal anneal at 800 K when the GB meets an open surface
at one end and is constrained by a wall of fixed atoms at the other.
In the simulation of the $\Sigma5\,(310)$ GB initiated with the normal
kite structure, a new structure nucleates at the surface and grows
inside the boundary. This structure can be identified as split kites,
whereas the untransformed part of the boundary continues to be composed
of normal kites. The extra atoms (\textasciitilde{}40\% of the $(310)$
plane) required for the transformation are supplied by diffusion from
the surface grove, a process which kinetically controls the transformation
rate. Eventually the new structure penetrates through the entire GB
and reaches the fixed region. When the fixed atoms on the left are
made dynamic (thus opening another surface), the entire GB reaches
the transformed state (Fig.~\ref{fig:310_800_trans}\emph{e}). 

The stability of the split-kite structure was tested by a series of
isothermal anneals. It remained stable and well-ordered above \textasciitilde{}800
K and until approximately 1150 K when it began to disorder and pre-melt.
Upon subsequent cooling from pre-melting temperatures to 1000 K, the
ordered split-kite structure reappeared inside the GB. This verifies
the thermodynamic stability of the split-kite structure and eliminates
any suspicion that its formation could be caused by the presence of
the surface grove (an additional proof comes from the point-defect
induced transformations in the absence of surfaces discussed below).
We were unable to demonstrate the reversibility of the transformation
back to normal kites by cooling the GB below 800 K because of the
slow (on the MD time scale) diffusion rates at low temperatures. However,
the zero-temperature results plotted in Fig.~\ref{fig:0K-structures}
establish that the normal kite structure has the lowest energy and
will be thermodynamically stable at sufficiently low temperatures.

Fig.~\ref{fig:310_800_trans}\emph{d} demonstrates that the two structures
of the $\Sigma5\,(310)$ GB can coexist and are separated by a line
defect with an atomic-scale cross-section. This allows us to consider
the two GB structures as two-dimensional (2D) phases separated a one-dimensional
(1D) phase boundary. The phase transformation is first order and is
kinetically controlled by GB diffusion. To our knowledge, this is
the first observation of coexistence of two GB phases separated by
a 1D phase boundary in atomistic simulations. It is also evident that
this phase boundary is associated with a GB step\cite{Hirth94}. The
exact nature of this 1D defect deserves a separate study in the future. 

Because the two GB structures have different interface stresses and
excess volumes, long-range elastic stresses could be generated around
the 1D boundary between them. In principle, such stresses could affect
the stability of the GB phases, e.g. via the image forces arising
due to the interaction of the stress field with boundary conditions.
In our case, however, this effect was negligible due to the large
size of the simulation block (about $12.5$ nm in the $y$-direction).
For additional verification, simulations were repeated with different
boundary conditions in the y-direction and the same GB phases were
invariably found.

\begin{figure}
\begin{centering}
\includegraphics[width=0.8\textwidth]{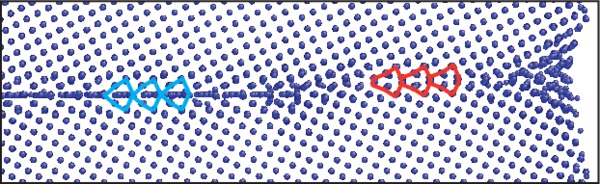}
\par\end{centering}

\caption{\textbf{High temperature transformation of the $\Sigma5\,(210)$ GB.}
MD simulation of the Cu $\Sigma5\,(210)$ GB at 800 K with an open
surface allowing variations in GB density. The boundary undergoes
a first order phase transformation from filled kites to split kites
nucleating and growing from the surface. \label{fig:S5(210)}}
\end{figure}

\begin{figure}
\begin{centering}
\includegraphics[height=0.8\textheight]{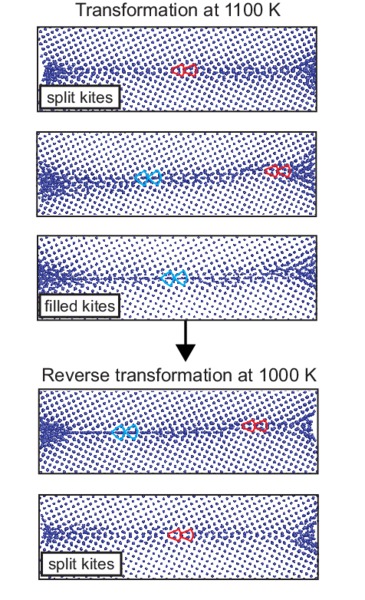}
\par\end{centering}

\caption{\textbf{Reversible GB phase transformations induced by temperature.}
$\Sigma5\,(210)$ GB terminated at two open surfaces. The initial
split-kite structure taken from a prior anneal at 1000 K transforms
into filled kites at 1100 K. Upon subsequent cooling and annealing
at 1000 K, it returns back to the split kites. Intermediate stages
show the two GB structures: one growing from the surface and the other
disappearing. It is concluded that the phase transition temperature
is between 1000 K and 1100 K. \label{fig:210_1100K_trans}}
\end{figure}

Similar transformations were found in the $\Sigma5\,(210)$ GB in
simulations starting from the filled-kite structure formed at \textasciitilde{}400
K. During isothermal anneals at temperatures below 1050 K, the GB
transforms into its thermodynamically stable phase at low temperatures:
the split kites. This new phase grows from the surface and eventually
penetrates through the entire boundary (cf. Fig.~\ref{fig:S5(210)}).
When the boundary with split kites was heated up to 1100 K, its structure
transformed to filled kites (Fig.~\ref{fig:210_1100K_trans}). In
this case, however, we were able to observe the reverse transformation.
Upon subsequent cooling down to 1000 K, the GB structure returned
to the split kites, demonstrating full reversibility of the phase
transformation. At temperatures close to \textit{T}\textsubscript{m}
this GB pre-melts by disordering of the filled-kite structure as observed
earlier\cite{Williams09}. It is interesting to note that the normal
kite structure traditionally attributed to this boundary never appears
as its thermodynamically stable phase. \bigskip{}

\noindent \textsf{\textbf{GB phase transformations caused by point
defects.}} During the phase transformations, the GB absorbs/ejects
large amounts of atoms from/to the surface. These absorption/ejection
processes may have important implications for properties of polycrystals
under extreme conditions. For example, in metals subject to radiation
by energetic particles, large numbers of vacancies and interstitials
are formed\cite{Grimes2008}. Interstitials are much more mobile than
vacancies and reach sinks first, whereas the remaining vacancies can
form clusters, stacking-fault tetrahedra and other defect complexes
degrading the material properties. GBs can act as effective sinks
of radiation-induced defects and can play a role in the damage healing.
Some of the mechanisms of defect-GB interactions have been studied
by MD simulations \cite{Bai26032010,Demkowicz08}. Our work suggests
that the absorption of point defects can strongly modify the GB structure,
causing structural transformations which, in turn, can greatly increase
the absorption capacity of the boundary.

\begin{figure}
\begin{centering}
\includegraphics[height=0.8\textheight]{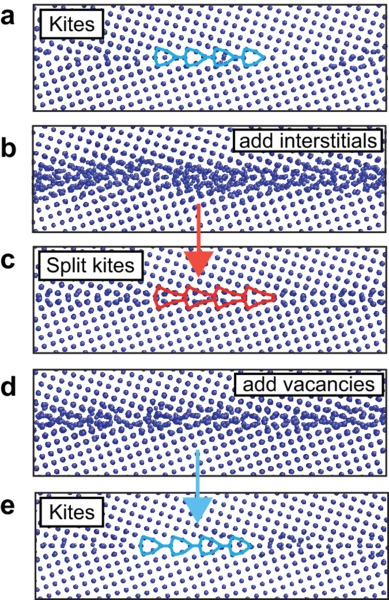}
\par\end{centering}

\caption{\textbf{Isothermal reversible GB phase transformations induced by
point defects.} GB phase transformations in the Cu $\Sigma5\,(310)$
GB induced by interstitials and vacancies in a simulation block with
periodic boundary conditions at $T=800$ K. After 80 interstitials
are introduced into a 10 $\textrm{\AA}$ thick layer containing the
GB, it transforms from the initial normal-kite structure (\textbf{a})
to a disordered state (\textbf{b}) and then to split kites (\textbf{c}).
After the subsequent introduction of 80 vacancies into the same GB
layer, the split-kite structure disorders (\textbf{d}) and then transforms
back to normal kites (\textbf{e}). The GB states (\textbf{a}) and
(\textbf{e}) are identical confirming that the transformation is fully
reversible. \label{fig:vav_inter_pbc}}
\end{figure}

The proposed effect is illustrated in Fig.~\ref{fig:vav_inter_pbc}.
The simulation was intentionally performed in a periodic simulation
block to exclude all sources/sinks of atoms other than the radiation
defects. The GB tested is Cu $\Sigma5\,(310)$ and its initial structure
consists of normal kites. This structure remains unchanged during
anneals when no atoms are added to or removed from the system. We
then introduce 80 interstitials into the GB region, which is equivalent
to 40\% of a $(310)$ atomic plane. This fraction of the plane matches
the local minimum of the GB energy corresponding to the split-kite
structure (cf.~Fig.~\ref{fig:0K-structures}\emph{a}). Upon anneal
at 800 K, the interstitials diffuse into the GB core and get absorbed
in it initially creating a disordered structure. As the anneal continues,
the GB eventually orders into the split-kite phase. To demonstrate
the reversibility of this radiation-induced transition, we then randomly
remove the same number of atoms from the GB region, which  simulates
the insertion of radiation-produced vacancies. After a period of disorder
the GB structure transforms back to the normal kites. Thus, in the
end of this phase transformation cycle the boundary returns to its
initial state having annihilated a large amount of radiation-induced
defects. A similar transformation cycle induced by absorption of point
defects was observed in the $\Sigma5\,(210)$ GB (Supplementary Figure
S4).\bigskip{}

\noindent \textsf{\textbf{Simulations of GB diffusion.}} As discussed
above, experimental measurements of GB diffusion in Cu bicrystals
containing the $\Sigma5\,(310)$ GB have been made by Budke et al.\cite{Budke1999}
and more recently by Divinsky et al\cite{Divinski2012}. The measurements
revealed a marked change in the slope of the Arrhenius plot of the
diffusivity at about 800-850 K\cite{Budke1999,Divinski2012}. To test
the prediction that this change is caused by a structural phase transformation
in this boundary, we performed MD simulations of the GB diffusivity
using the methodology proposed in Ref.\cite{Suzuki05a,Frolov09c,Frolov2012c}.
The simulations were performed separately for the normal-kite and
split-kite structures. Recall that the normal kites are stable at
low temperatures and transform to split kites at temperatures 800
K and higher. 

\begin{figure}
\noindent \begin{centering}
\includegraphics[width=0.7\textwidth]{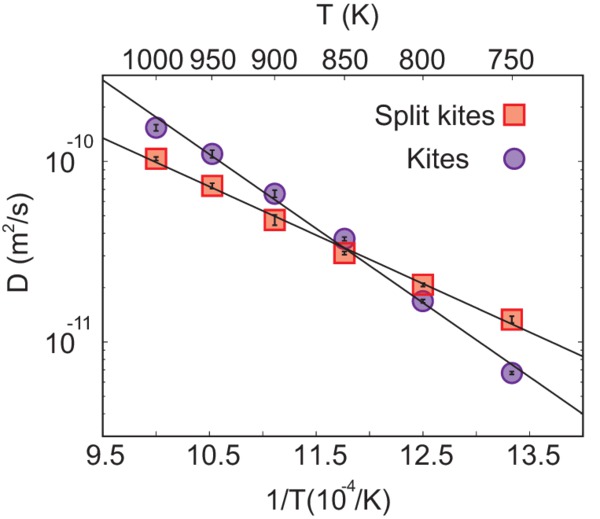}
\par\end{centering}

\caption{\textbf{Arrhenius diagram of the diffusion coefficient ($D$) calculated
for two structures $\Sigma5\,(310)$ boundary in Cu.} Note the difference
in the activation energy between the two GB structures. \label{fig:GB-diffus}}
\end{figure}

The results of the diffusion calculations are shown on the Arrhenius
diagram in Fig.~\ref{fig:GB-diffus}. Note that the two GB structures
have different diffusivities as well as activation energies. Furthermore,
it is observed that the high-temperature split-kite phase has a lower
activation energy (smaller slope) than the low-temperature normal-kite
phase (larger slope). This trend closely correlates with the experimental
measurements\cite{Budke1999,Divinski2012}, as does the approximate
temperature at which the change in the slope of the Arrhenius diagram
is expected. This agreement strongly suggests that the change in the
GB diffusivity measured in the experiments\cite{Budke1999,Divinski2012}
was caused by the phase transformation in this boundary. Further details
of the diffusion simulations will be published elsewhere. \bigskip{}

\noindent \textsf{\textbf{\Large Discussion}}{\Large \par}

In this work we demostrate by atomistic simulations that high-angle
metallic GBs can have multiple stable or metastable phases with different
atomic structures and first-order transitions between them. For the
particular symmetrical tilt $\Sigma5$ GBs studied here, we found
three different phases. The normal kite structure predicted to be
the 0 K ground state of the $\Sigma5\,(310)$ GB is consistent with
room-temperature HRTEM observations\cite{Duscher04}. The remaining
two structures have not been seen by HRTEM yet but could have manifested
themselves through sudden changes in temperature dependencies of GB
properties such as diffusivity\cite{Divinski2012,Budke1999}. In fact,
our simulations have shown that the GB phases found here are indeed
characterized by different GB diffusion coefficients and that the
slopes of their Arrhenius plots correlate with the experiment\cite{Divinski2012,Budke1999}.
It should be recognized, however, that the present diffusion calculations
can be compared with the experiment\cite{Divinski2012,Budke1999}
only in terms of trends but not quantitatively. While our diffusion
calculations were performed for Cu self-diffusion, the experimental
measurements were conducted for impurity diffusion of Ag in Cu. The
latter process is accompanied by segregation of the diffusing atoms,
which may potentially enhance the effect of the structural transformations
on GB diffusion or even alter the GB phases and/or their transformations
temperatures. The potential effects of solute segregation lie outside
the scope of this paper.

All three GB phases found in this work were reproduced in several
FCC metals modeled with different atomistic potentials, suggesting
that they represent a generic feature of FCC crystals. More general
GBs may feature a larger variety of phases. The existence of GB phases
in metals is likely to be a common phenomenon, which was previously
overlooked due to the overly constrained computer simulation methodology
and challanges inherent in experimental atomic-level observations
of GB structures at elevated temperatures.

The finding of first-order transitions in elemental high-angle GBs
is important as it provides a new testing ground for fundamental theories
of GB thermodynamics and GB phase transitions\cite{Cahn82a}. Future
calculations of the exact transformation temperatures, transformation
enthalpies as well as the excess thermodynamics properties of the
1D phase boundaries would be of significant fundamental interest.
In the future this study could be extended to multicomponent systems
which may additionally exhibit segregation-induced structural transformations
with rather complex phase diagrams. In addition to the fundamental
questions raised by this work, the present simulations suggest a possible
novel mechanism of radiation damage healing in which a GB absorbs
large amounts of point defects by transformation between different
phases. It is believed that the same mechanism of defect absorption
can operate under other extreme conditions involving high non-equilibrium
concentrations of point defects, such as severe plastic deformation,
diffusional creep, and rapid grain growth in nano-crystalline materials.

\bigskip{}

\noindent \textsf{\textbf{\large Methods}}{\large \par}

\noindent \textsf{\textbf{GB structure calculations at 0 K. }}For
each boundary, its energy was minimized for several different cross-sections
of the periodic simulation block. A cross-section is defined by $n\times m$,
where $n$ and $m$ are the numbers of lattice periods in the directions
parallel and normal to the tilt axis, respectively. For the $\Sigma5\,(210)$
GB we used the following cross-sections: $4\times2$, $6\times1$,
$2\times1$, $7\times2$, $9\times2$, $11\times2$, $13\times2$
and $15\times2$. For the $\Sigma5\,(310)$ GB the cross-sections
were $10\times2$, $8\times2$, $9\times6$, $9\times2$, $7\times2$
and $13\times2$. 

For each GB with a given cross-section, a number of trial configurations
were generated and then minimized as follows, with the lowest minimized
energy reported. First, relative translations of the two grains by
vectors within a primitive cell of the displacement shift complete
(DSC) lattice and all possible, non-equivalent, placements of the
boundary plane in the direction normal to the boundary were generated.
(These different placements are equivalent to relative translations
by full DSC vectors.) For example, for the $7\times2$ searches, $3\times3\times3$
vectors within the DSC primitive cell were used, and the total number
of configurations generated at this stage, including the different
placements of the boundary plane, was 225. These will be referred
to as base configurations below. For some searches more points within
the DSC primitive cell were used. 

For each base configuration, all atoms within 15 $\textrm{\AA}$ of
the nominal boundary plane were moved 0.001 $\textrm{\AA}$ in a (pseudo)
random direction. Atoms that were very near each other were then removed
to generate the configurations to be minimized. At each step of the
removal process, the closest two atoms were both removed and an atom
put back at the average position of the two that were removed. The
process continued until no pair of atoms was within about 1/3 of the
nearest neighbor distance. This was the first trial configuration
to be minimized. After that, one atom was removed at a time (from
the non-minimized configuration) to generate a configuration to be
minimized. This was done until no pair of atoms was within about $0.99$
of the nearest neighbor distance. In total, about 16,500 configurations
were minimized in a $7\times2$ search for the $\Sigma5\,(310)$ GB
and about 24,000 for the $\Sigma5\,(210)$ GB. The total number $N$
of atoms removed could be larger than the number $N_{0}$ of atoms
in a perfect plane parallel to the GB. In such cases, the GB density
was determined by dividing $N$ by $N_{0}$ and expressing the remainder
as a fraction of $N_{0}$. \bigskip{}

\noindent \textsf{\textbf{MD simulations at finite temperatures.}}
The MD simulations were performed in the canonical (NVT) ensemble
with a Nosé-Hoover thermostat implemented in the ITAP molecular dynamics
(IMD) program\cite{StadlerMT97}. The two symmetrical tilt GBs, $\Sigma5\,(310)\,[001]$
and $\Sigma5\,(210)\,[001]$, were created by joining two crystals
rotated relative to each other around the {[}001{]} direction by the
angle of 36.87º and 53.13º, respectively. The $[001]$ tilt axis of
the GB was aligned parallel to the $x$ direction with the GB normal
parallel to the $y$ direction (Supplementary Figure S3). The system
typically contained 150,000 to 250,000 atoms. Smaller blocks, containing
about 50,000 atoms with periodic boundary conditions parallel to the
GB plane, were used used to compute the GB thickness, observe the
GB premelting and simulate the GB transformations induced by vacancies
and interstitials. In the simulations involving large blocks, the
periodic boundary condition was kept in the $x$ direction, whereas
in the $z$ direction the GB terminated at an open surface on one
side and at a layer of fixed atoms on the other side. In some simulations,
the GB terminated at open surfaces on both sides. 

The isothermal anneals of the GBs were performed at temperatures from
0 K to $0.98$\textit{ T\textsubscript{\textit{m}}}. The boundary
conditions were either fully periodic in the directions parallel to
the grain boundary plane or non-periodic when the GB was terminated
at an open surface (Supplementary Figure S3). Additional simulations
were conducted for the following FCC metals modeled with EAM potentials:
Cu\cite{Mishin01,VoterCu}, Ag\cite{Williams06}, Au\cite{Foiles86}
and Ni\cite{Foiles06a}. The GB atomic density changes were evaluated
by comparing the number of atoms in a simulation block containing
a given GB with the number of atoms in the initial block of the same
cross-section containing an integer number of complete atomic planes.
Relaxed GB structures were used to calculate the excess GB properties
such as GB energy, volume and stress \cite{Frolov2012a,Frolov2012b}.

The calculations of the GB diffusion coefficients employed the methodology
which was described in detail in previous work\cite{Suzuki05a,Frolov09c,Frolov2012c}
and will not be repeated here.

\bigskip{}

\noindent \textsf{\textbf{MD simulations GB transformations induced
by point defects.}} In such simulations, the GB initially had the
normal kite structure which was first annealed at 800 K for several
ns. Then, 80 and 132 interstitials were randomly introduced inside
a 1 nm thick layer containing the $\Sigma5\,(310)$ and $\Sigma5\,(210)$
GBs, respectively. These numbers of atoms correspond to \textasciitilde{}40\%
and \textasciitilde{}50\% of the $(310)$ and $(210)$ planes, respectively.
After the atoms were inserted, the isothermal anneal was continued
at 800 K. The initial kite (respectively, filled kite) structures
quickly disappeared and significant atomic rearrangements took place
in the boundary layer. However, within 1 to 100 ps of the simulation
time the boundaries re-ordered into the split-kite structure. 

The obtained split-kite structures of both GBs were then taken as
initial states to introduce vacancies in order to demonstrate transformations
into metastable phases. The vacancies were introduced by randomly
removing atoms within a 1 nm thick layer containing the GB. The number
of vacancies was equal to the number of interstitials introduced previously.
After 1 to 100 ps of the isothermal anneal at 800 K and a period of
disorder, we observed a transformation into the normal kite structure
in the $\Sigma5\,(310)$ GB and the filled-kite structure in the $\Sigma5\,(210)$
GB.

\section*{Acknowledgements}

T.F. was funded by the Miller Institute. Y.M. was supported by the
U.S. Department of Energy, the Physical Behavior of Materials Program,
through Grant No. DE-FG02-01ER45871. D.L.O. was supported by the U.S.
Department of Energy through Grant No. DE-AC02-05CH11231.

\section*{Author contributions}

T.F. performed the MD simulations, D.L.O. performed the 0 K GB structure
calculations, M.A. and Y.M. directed this work. All co-authors participated
in the interpretation of the results and manuscript preparation.

\section*{Competing financial interests}

The authors declare that they have no competing financial interests.


\newpage{}\clearpage{}

\bigskip{}

\begin{table}
\begin{centering}
\begin{tabular}{|c|c|c|c|c|c|c|}
\hline 
Structure & Cross-section & Fraction of plane & $[U]_{N}$, J/m$^{2}$ & $[V]_{N}$, Å & $\tau_{11}^{N}$, J/m$^{2}$ & $\tau_{22}^{N}$, J/m$^{2}$\tabularnewline
\hline 
\multicolumn{7}{|c|}{$\Sigma5\,(310)$ GB}\tabularnewline
\hline 
Normal kites (0 K) & $1\times1$ & $0$ & 0.9047 & 0.316 & 1.305 & 1.774\tabularnewline
\hline 
Split kites (0 K) & $10\times2$ & $2/5$  & 0.911 & 0.233 & 0.493 & 0.0465\tabularnewline
\hline 
Split kites (MD) & $18\times6$ $^{*}$ & $0.37$ & 0.920 & 0.245 & 0.517 & 0.0971\tabularnewline
\hline 
\multicolumn{7}{|c|}{$\Sigma5\,(210)$ GB}\tabularnewline
\hline 
Normal kites (0 K) & $1\times1$ & $0$ & 0.951 & 0.322 & 1.174 & 1.491\tabularnewline
\hline 
Split kites (0 K) & $15\times2$ & $7/15$  & 0.936 & 0.172 & 0.297 & -2.12\tabularnewline
\hline 
Split kites (MD) & $18\times38$ $^{*}$ & $0.46$ & 0.98 & 0.23 & 0.446 & -1.276\tabularnewline
\hline 
Filled kites (0 K) & $7\times2$ & $6/7$ & 0.953 & 0.301 & 1.042 & 2.299\tabularnewline
\hline 
\end{tabular}
\par\end{centering}

\caption{\textbf{Characteristics of different structures of the $\Sigma5$
GBs in Cu modeled with the EAM1}\cite{Mishin01} \textbf{potential.}
Note that the periodic unit cells of the structures are different
and can be significantly larger than the repeat unit of the lattice.
$^{*}$The MD unit cell represents the size of the simulation block
for which the GB properties were calculated and is larger than the
smallest periodic unit.\label{tab:310}}
\end{table}

\bigskip{}

\bigskip{}


\newpage{}\clearpage{}\setcounter{page}{1}

\title{\textsf{\textbf{\large Structural phase transformations in metallic
grain boundaries}}\bigskip{}
}

\author{Timofey Frolov$^{1\star}$, \enskip{}David L. Olmsted$^{1}$,\enskip{}Mark
Asta$^{1}$ and Yuri Mishin$^{2}$}

\lyxaddress{$^{1}$ Department of Materials Science and Engineering, University
of California, Berkeley, California 94720, USA}

\lyxaddress{$^{2}$ School of Physics, Astronomy and Computational Sciences,
George Mason University, Virginia, 22030, USA}

$^{\star}$e-mail: timfrol@berkeley.edu

\bigskip{}

\bigskip{}

\noindent \textbf{Supplementary Materials include:}

\noindent \textbf{Supplementary Figures S1, S2, S3 and S4}\newpage{}

\begin{figure}
\begin{centering}
\includegraphics[width=0.7\textwidth]{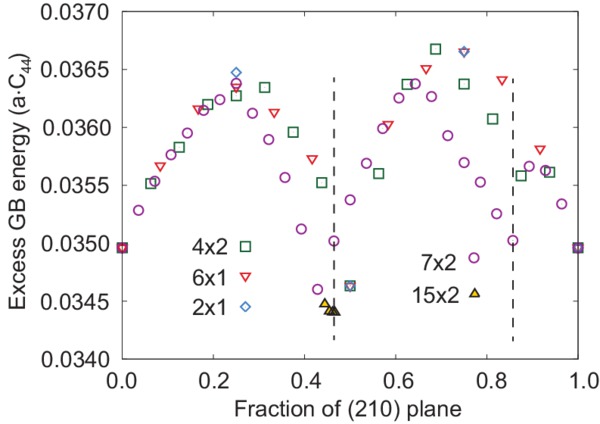}
\par\end{centering}

\bigskip{}

\noindent \begin{centering}
\textbf{Supplementary Figure S1. }
\par\end{centering}

\textbf{Examples of GB energy searches.} The examples are for the
$\Sigma5\,(210)$ GB in Cu modeled with the EAM1 potential\cite{Mishin01}
with different GB cross-sections. The GB energy is measured in the
units of $ac_{44}$, where $a$ is the lattice constant and $c_{44}$
is the elastic constant. Note that the individual searches exhibit
sharp minima at different fractions of the $(210)$ plane. \label{fig:Evsfrac_210_searches} 
\end{figure}

\newpage{}\clearpage{}
\begin{figure}
\begin{centering}
\includegraphics[height=0.7\textheight]{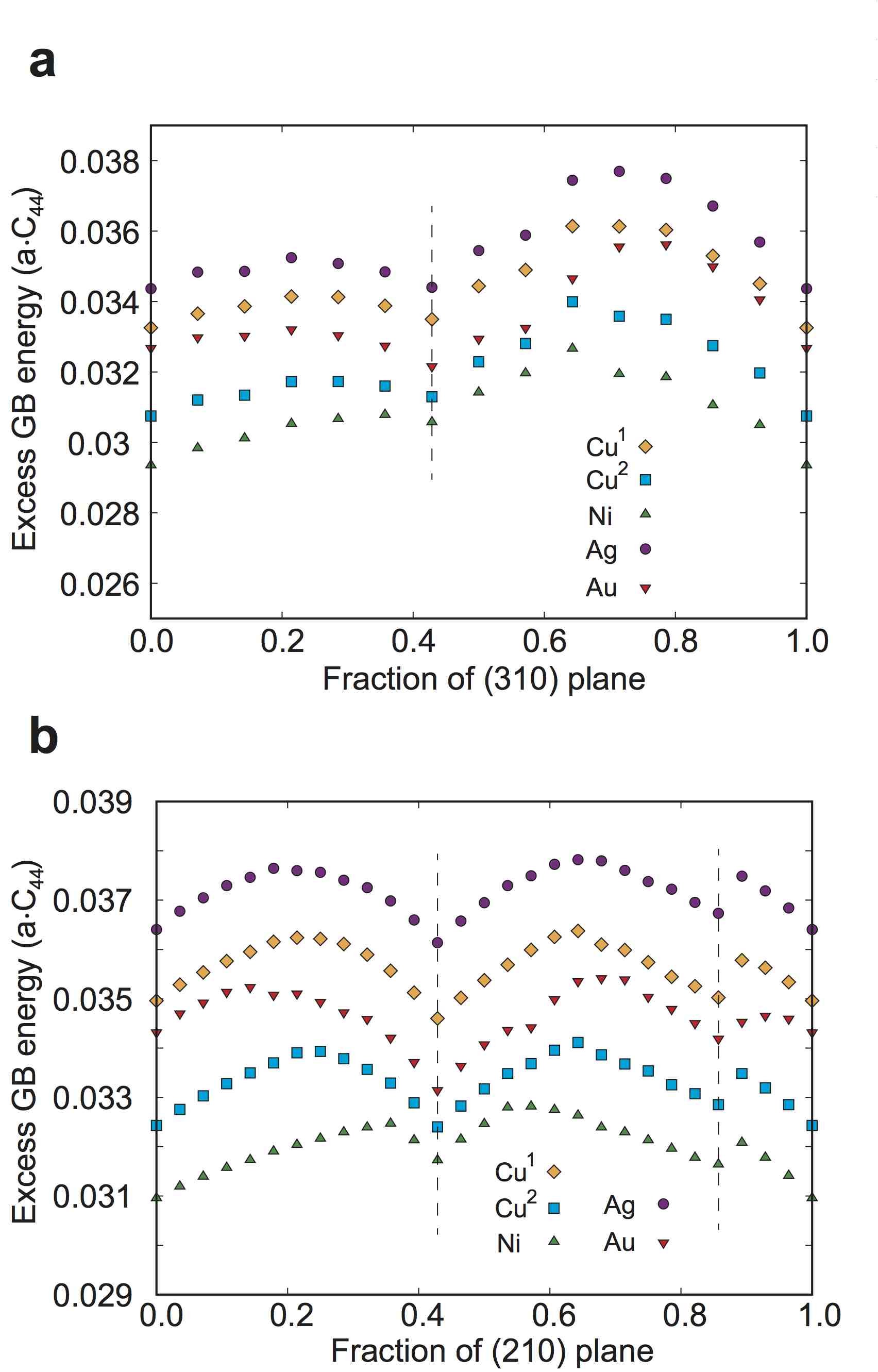}
\par\end{centering}

\bigskip{}

\noindent \begin{centering}
\textbf{Supplementary Figure S2. }
\par\end{centering}

\textbf{GB energy calculations for different FCC metals.} The metals
are modeled with different EAM potentials independently developed
by different groups over the past 25 years: EAM1-Cu\cite{Mishin01},
EAM2-Cu\cite{Mishin01}, Ag\cite{Williams06}, Au\cite{Foiles86}
and Ni\cite{Foiles06a}. To facilitate comparison, the GB energy is
measured in the units of $ac_{44}$, where $a$ is the lattice constant
and $c_{44}$ is the elastic constant. (a) $\Sigma5\,(310)$ GB. (b)
$\Sigma5\,(210)$ GB. All calculations used the same $7\times2$ cross-section
measured in lattice repeat distances of the bulk crystal. Note that
all of these potentials identify the normal kite, split kite and filled
kite configurations as local minima of the GB energy. However, the
relative energies of these GB structures are different for different
metals. This demonstrates that the low-energy GB structures identified
in this work are not specific to the particular Cu potential but are
likely to represent a generic feature of several FCC metals. \label{fig:Evsfrac7x2pots}
\end{figure}

\newpage{}

\clearpage{}
\begin{figure}
\begin{centering}
\includegraphics[width=0.9\textwidth]{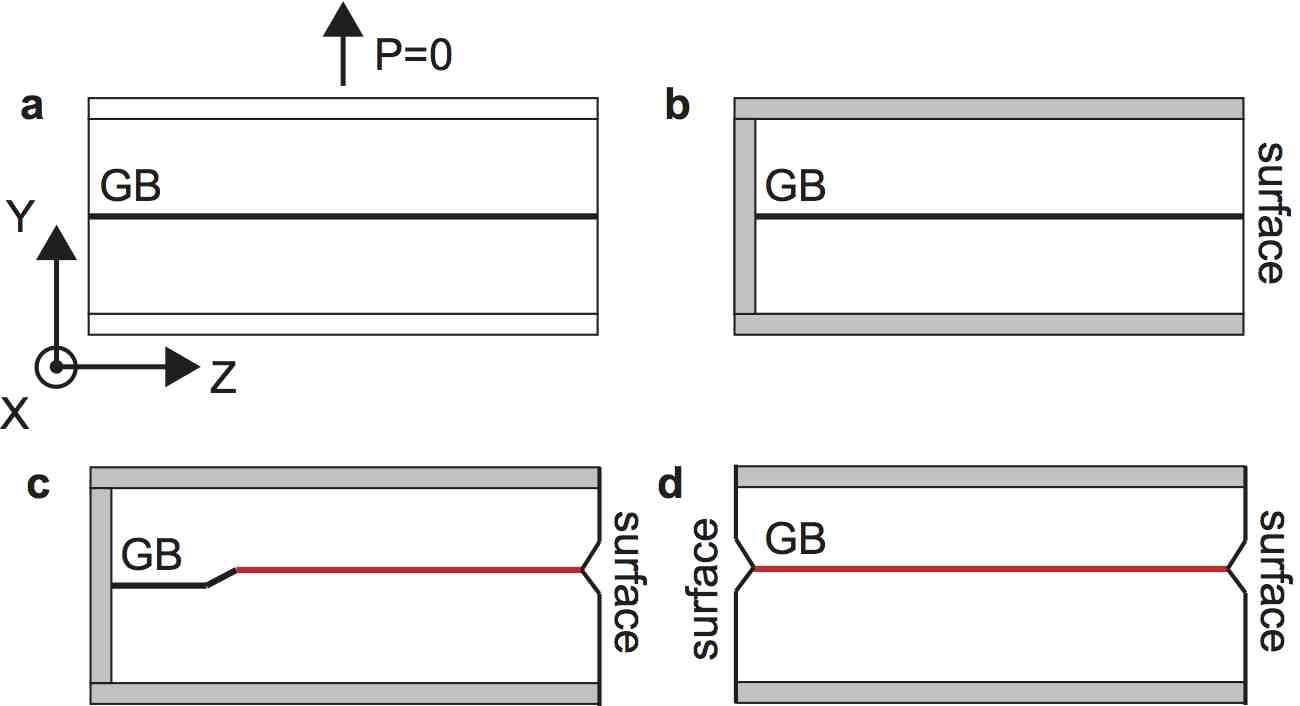}
\par\end{centering}

\begin{centering}
\bigskip{}

\par\end{centering}

\noindent \begin{centering}
\textbf{Supplementary Figure S3. }
\par\end{centering}

\textbf{Schematic illustration of the simulation block with a GB phase
transformation during isothermal anneals.} (a) Simulation block with
periodic boundary conditions in the $x$ and $z$ directions. In the
$y$ direction the block terminates at free surfaces. (b) Initial
structure of the simulation block with two fixed regions in the $y$
direction and one the $z$ direction. One side of the GB in the $z$
direction is terminated at the open surface. (c) GB transformation
during an isothermal anneal. (d) The GB transformation is completed
after a second surface has been created in the $z$ direction. \label{fig:cartoon}
\end{figure}

\newpage{}

\clearpage{}
\begin{figure}
\begin{centering}
\includegraphics[height=0.8\textheight]{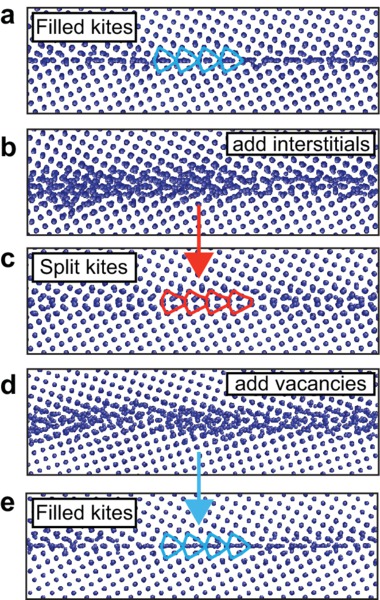}\bigskip{}

\par\end{centering}

\noindent \begin{centering}
\textbf{Supplementary Figure S4. }
\par\end{centering}

\textbf{GB phase transformations induced by point defects. }Phase
transformations in the $\Sigma5\,(210)$ GB induced by interstitials
and vacancies in a simulation block with periodic boundary conditions
at $T=800$ K. The GB transforms from filled kites (a) to split kites
(c) after 132 interstitials were introduced in a 1 nm thick layer
containing the GB. The GB transforms from the split kites (c) back
to filled kites (e) after 132 vacancies were introduced in same layer.
(b) and (d) demonstrate that significant atomic rearrangements take
place during the transformations. Note that the initial (a) and final
(e) structures are identical. This metastable filled-kite structure
does not spontaneously transform into the more stable split-kite structure
due to the absence of sources and sinks of atoms in the periodic simulation
block.\label{fig:210_vac_inter}
\end{figure}

\end{document}